# Some Implications of Two Forms of the Generalized Uncertainty Principle

**Mohammed M. Khalil** *

*Department of Electrical Engineering, Alexandria University, Alexandria 12544, Egypt*

Various theories of quantum gravity predict the existence of a minimum length scale, which leads to the modification of the standard uncertainty principle to the Generalized Uncertainty Principle (GUP). In this paper, we study two forms of the GUP and calculate their implications on the energy of the harmonic oscillator and the Hydrogen atom more accurately than previous studies. In addition, we show how the GUP modifies the Lorentz force law and the time-energy uncertainty principle.

## 1. Introduction

Developing a theory of quantum gravity is currently one of the main challenges in theoretical physics. Various approaches predict the existence of a minimum length scale [1, 2] that leads to the modification of the Heisenberg Uncertainty Principle:

$$\Delta x\ \Delta p \geq \hbar / 2, \tag{1}$$

to the Generalized Uncertainty Principle (GUP) [3, 4]:

$$\Delta x\ \Delta p \geq \frac{\hbar}{2}\left(1+\beta\left(\Delta p\right)^2+\zeta\right), \tag{2}$$

where $\beta = \beta_0 l_P{}^2 / \hbar^2$ , $\beta_0$ is a dimensionless constant usually assumed to be of order unity, $l_P \equiv \sqrt{\hbar G / c^3}$ is the Planck length $l_P \simeq 1.616\times 10^{-35} m$ , and $\zeta$ may depend on $\langle p \rangle$ but not on $\Delta p$ . The second term on the RHS above is important at very high energies/ small length scales (i.e. $\Delta x \sim l_P$ ).

In this article, we study two forms of the GUP. The first (GUP1) [5, 6] is:

$$\Delta x_i\ \Delta p_i \geq \frac{\hbar}{2}\left[1+\beta\left(\left(\Delta p\right)^2+\langle p \rangle^2\right)+2\beta\left(\left(\Delta p_i\right)^2+\langle p_i \rangle^2\right)\right], \tag{3}$$

which follows from the modified commutation relation [6]:

$$\left[x_i, p_j\right] = i\hbar\left(\delta_{ij}+\beta\left(p^2\delta_{ij}+2p_i p_j\right)\right). \tag{4}$$

The second (GUP2) [7, 8] is:

$$\Delta x\ \Delta p \geq \frac{\hbar}{2}\left[1-2\alpha\langle p \rangle+4\alpha^2\left(\left(\Delta p\right)^2+\langle p \rangle^2\right)\right]. \tag{5}$$

which follows from the proposed modified commutation relation [7]:

$$\left[x_i, p_j\right] = i\hbar\left(\delta_{ij}-\alpha\left(p\delta_{ij}+\frac{p_i p_j}{p}\right)+\alpha^2\left(p^2\delta_{ij}+3p_i p_j\right)\right), \tag{6}$$

where $\alpha = \alpha_0 l_P / \hbar = \alpha_0 / M_P c$ , $\alpha_0$ is a constant usually assumed to be of order unity. In addition to a minimum measurable length, GUP2 implies a maximum measurable momentum.

The commutation relation (4) admits the following representation in position space [9, 10]:

$$x_i = x_{0i}, \qquad p_i = p_{0i}\left(1+\beta p_0^2\right), \tag{7}$$

where $x_{0i}, p_{0i}$ satisfy the canonical commutation relation $\left[x_{0i}, p_{0j}\right] = i\hbar\delta_{ij}$. This definition modifies any Hamiltonian near the Planck scale to [9, 10]:

$$H = \frac{p_0^2}{2m}+V(r)+\frac{\beta}{m}p_0^4+\frac{\beta^2}{2m}p_0^6. \tag{8}$$

Similarly, (6) admits the definition [7, 8]:

$$x_i = x_{0i}, \qquad p_i = p_{0i}\left(1-\alpha p_0+2\alpha^2 p_0{}^2\right), \tag{9}$$

---

* e-mail: moh.m.khalil@gmail.com

leading to the perturbed Hamiltonian:

$$H = \frac{p_0^2}{2m} + V\left(r\right) - \frac{\alpha}{m}p_0^3 + \frac{5\alpha^2}{2m}p_0^4 - \frac{2\alpha^3}{m}p_0^5 + \frac{2\alpha^4}{m}p_0^6 . \tag{10}$$

The aim of this article is to study the impact of GUP1 and GUP2 on the energy of the harmonic oscillator and Hydrogen atom more accurately than previous studies. In addition, we show how the GUP modifies the Lorentz force law and the time-energy uncertainty principle.

## 2. Harmonic Oscillator

The harmonic oscillator is a good model for many systems, so it is important to calculate its energy accurately to compare it with future experiments. Recently a quantum optics experiment was proposed [11] to probe the commutation relation of a mechanical oscillator with mass close to the Planck mass.

The effect of GUP1 on the eigenvalues of the harmonic oscillator was calculated exactly in [12]. The effect of GUP2 was considered in [8] to first and second order for the ground energy only. In this section, we consider first and second order corrections to all energy levels for both GUPs to compare them, and we use the ladder operator method, which is simpler than the other methods.

**GUP1- first order:**

The momentum $p_0$ can be expressed using the ladder operators [13, P.49] as:

$$p_0 = i\sqrt{\frac{\hbar m\omega}{2}}\left(a^\dagger - a\right), \tag{11}$$

where $a^\dagger$ is the raising operator: $a^\dagger\left|n\right\rangle = \sqrt{n+1}\left|n+1\right\rangle$, and $a$ is the lowering operator: $a\left|n\right\rangle = \sqrt{n}\left|n-1\right\rangle$. Thus, the change in energy to first order due to $H' = \beta p_0^4/m + \beta^2 p_0^6/2m$ is:

$$\Delta E_{n(GUP1)}^{(1)} = \left\langle n\right|H'\left|n\right\rangle = \frac{\beta}{m}\left(\frac{\hbar m\omega}{2}\right)^2\left\langle n\right|\left(a^\dagger - a\right)^4\left|n\right\rangle - \frac{\beta^2}{2m}\left(\frac{\hbar m\omega}{2}\right)^3\left\langle n\right|\left(a^\dagger - a\right)^6\left|n\right\rangle . \tag{12}$$

Applying the raising and lowering operators, and simplifying:

$$\Delta E_{n(GUP1)}^{(1)} = \frac{3\beta_0 l_P^2}{4}m\omega^2\left(2n^2+2n+1\right) + \frac{5\beta_0{}^2 l_P^4 m^2\omega^3}{16\hbar}\left(4n^3+6n^2+8n+3\right). \tag{13}$$

Therefore, the relative change in energy is:

$$\frac{\Delta E_{n(GUP1)}^{(1)}}{E_n} = \frac{3\beta_0 l_P^2 m\omega}{4\hbar}\frac{\left(2n^2+2n+1\right)}{n+1/2} + \frac{5\beta_0{}^2 l_P^4 m^2\omega^2}{16\hbar^2}\frac{\left(4n^3+6n^2+8n+3\right)}{n+1/2}. \tag{14}$$

The first term in (13) differs from that derived in [12] by a factor of three, because instead of the commutation relation (4) they use the relation $\left[x,p\right] = i\hbar\left(1+\beta p^2\right)$.

**GUP1- second order:**

The second order correction can be calculated using second order perturbation theory [13, P.256]:

$$\Delta E_{n(GUP1)}^{(2)} = \sum_{m\neq n}\frac{\left|\left\langle m\right|H'\left|n\right\rangle\right|^2}{E_n^{(0)} - E_m^{(0)}}, \qquad H' = \frac{\beta}{m}p_0^4, \tag{15}$$

Expanding and neglecting terms with equal number of $a$ and $a^\dagger$:

$$\begin{aligned}\left\langle m\right|H'\left|n\right\rangle = &\frac{\beta}{m}\left(\frac{\hbar m\omega}{2}\right)^2 \\ &\left\langle m\right|\left(a^\dagger a^\dagger a^\dagger a^\dagger - a^\dagger a^\dagger a^\dagger a - a^\dagger a^\dagger a a^\dagger - a^\dagger a a^\dagger a^\dagger - a^\dagger aaa - aa^\dagger a^\dagger a - aa^\dagger aa - aaa^\dagger a - aaaa^\dagger + aaaa\right)\left|n\right\rangle,\end{aligned} \tag{16}$$

Applying the raising and lowering operators:

$$\left\langle m\right|H'\left|n\right\rangle = \frac{\beta}{m}\left(\frac{\hbar m\omega}{2}\right)^2\begin{pmatrix}\sqrt{\left(n+1\right)\left(n+2\right)\left(n+3\right)\left(n+4\right)}\delta_{m,n+4} - \left(4n+6\right)\sqrt{\left(n+1\right)\left(n+2\right)}\delta_{m,n+2} \\ -\left(4n-2\right)\sqrt{n\left(n-1\right)}\delta_{m,n-2} + \sqrt{n\left(n-1\right)\left(n-2\right)\left(n-3\right)}\delta_{m,n-4}\end{pmatrix} \tag{17}$$

Because of the delta functions and the orthogonality of the Eigenfunctions, squaring the above expression means squaring each term individually. After simplifying and dividing by $E_n$:

$$\frac{\Delta E_{n(GUP1)}^{(2)}}{E_n} = \frac{-m^2\omega^2 l_P^4\beta_0^2}{8\hbar^2}\frac{\left(34n^3+51n^2+59n+21\right)}{n+1/2}. \tag{18}$$

**GUP2- first order:**

For GUP2, $H' = \frac{-\alpha}{m} p_0^3 + \frac{5\alpha^2}{2m} p_0^4 - \frac{2\alpha^3}{m} p_0^5 + \frac{2\alpha^4}{m} p_0^6$ .The $p_0^3$ and $p_0^5$ terms do not contribute to first order because they are odd functions. The first order correction for the $p_0^4$ and $p_0^6$ terms is the same as (14) with $\beta \to 5\alpha^2/2$ and $\beta^2 \to 4\alpha^4$:

$$\frac{\Delta E^{(1)}_{n(GUP2)p_0^4}}{E_n} = \frac{15 l_P^2 \alpha_0^2 m\omega}{8\hbar} \frac{\left(2n^2+2n+1\right)}{n+1/2} + \frac{5\alpha_0^4 l_P^4 m^2\omega^2}{4\hbar^2} \frac{\left(4n^3+6n^2+8n+3\right)}{n+1/2}, \tag{19}$$

which agrees with the expression derived in [8] when $n=0$.

**GUP2- second order:**

The second order correction for the $p_0^3$ term can be calculated using the same method that led to (18):

$$\Delta E^{(2)}_{n(GUP2)p_0^3} = \sum_{m\neq n} \frac{\left|\langle m|H'|n\rangle\right|^2}{E_n^{(0)} - E_m^{(0)}}, \qquad H' = \frac{-\alpha}{m} p_0^3 \tag{20}$$

$$\langle m|H'|n\rangle = \frac{i\alpha}{m}\left(\frac{\hbar m\omega}{2}\right)^{\frac{3}{2}} \left(\sqrt{(n+1)(n+2)(n+3)}\,\delta_{m,n+3} - 3(n+1)\sqrt{n+1}\,\delta_{m,n+1} + 3n\sqrt{n}\,\delta_{m,n-1} - \sqrt{n(n-1)(n-2)}\,\delta_{m,n-3}\right) \tag{21}$$

Squaring and substituting in (20):

$$\Delta E^{(2)}_{n(GUP2)p_0^3} = \frac{\alpha^2}{m^2}\left(\frac{\hbar m\omega}{2}\right)^3 \left[\frac{(n+1)(n+2)(n+3)}{-3\hbar\omega} + \frac{9(n+1)^3}{-\hbar\omega} + \frac{9n^3}{\hbar\omega} + \frac{n(n-1)(n-2)}{3\hbar\omega}\right]. \tag{22}$$

Simplifying and dividing by $E_n$ :

$$\frac{\Delta E^{(2)}_{n(GUP2)p_0^3}}{E_n} = \frac{-m\omega l_P^2 \alpha_0^2}{8\hbar} \frac{\left(30n^2+30n+11\right)}{n+1/2}, \tag{23}$$

which agrees with the expression derived in [8] when $n=0$.

The second order correction for the $p_0^4$ term is the same as Eq. (18) with $\beta \to 5\alpha^2/2$ :

$$\frac{\Delta E^{(2)}_{n(GUP2)p_0^4}}{E_n} = \frac{-25 m^2\omega^2 l_P^4 \alpha_0^4}{32\hbar^2} \frac{\left(34n^3+51n^2+59n+21\right)}{\left(n+1/2\right)}. \tag{24}$$

Adding (14) and (18) we get for GUP1:

$$\frac{\Delta E_{n(GUP1)}}{E_n} = \frac{3\beta_0 l_P^2 m\omega}{4\hbar} \frac{\left(2n^2+2n+1\right)}{n+1/2} - \frac{3\beta_0^2 l_P^4 m^2\omega^2}{16\hbar^2} \frac{\left(16n^3+24n^2+26n+9\right)}{n+1/2} \tag{25}$$

Adding (19), (23) and (24) we get for GUP2:

$$\frac{\Delta E_{n(GUP2)}}{E_n} = \frac{m\omega l_P^2 \alpha_0^2}{2\hbar} \frac{1}{n+1/2} - \frac{15 m^2\omega^2 l_P^4 \alpha_0^4}{32\hbar^2} \frac{\left(46n^3+69n^2+77n+27\right)}{n+1/2} \tag{26}$$

It is interesting to note that to $O(\alpha^2)$ , the effect of GUP2 is to add a constant shift to all energy levels.

To compare (25) and (26) with experiment, consider an ion in a Penning trap; its motion is effectively a one-dimensional harmonic oscillator [14]. The accuracy of mass determination increases linearly with charge, so let us suppose it is possible to use completely ionized lead atoms, which have an atomic number of 82. Suppose that the magnetic field in the Penning trap is $B = 10\ T$. The cyclotron frequency is $\omega_c = qB/m$ ; substituting the value of $m\omega_c \simeq 820e$ in (25) and (26) we get the results shown in table 1 for different $n$ .

| **n** | $\Delta E_{n(GUP1)}/E_n$ | $\Delta E_{n(GUP2)}/E_n$ |
|---|---|---|
| **0** | $4.9\times10^{-52}\beta_0 - 3.6\times10^{-103}\beta_0^2$ | $3.2\times10^{-52}\alpha_0^2 - 2.7\times10^{-102}\alpha_0^4$ |
| **2** | $1.3\times10^{-51}\beta_0 - 2.3\times10^{-102}\beta_0^2$ | $6.5\times10^{-53}\alpha_0^2 - 1.6\times10^{-101}\alpha_0^4$ |
| **5** | $2.7\times10^{-51}\beta_0 - 9.9\times10^{-102}\beta_0^2$ | $3.0\times10^{-53}\alpha_0^2 - 7.1\times10^{-101}\alpha_0^4$ |
| **10** | $5.1\times10^{-51}\beta_0 - 3.5\times10^{-101}\beta_0^2$ | $1.5\times10^{-53}\alpha_0^2 - 2.5\times10^{-100}\alpha_0^4$ |
| **100** | $4.9\times10^{-50}\beta_0 - 3.2\times10^{-99}\beta_0^2$ | $1.6\times10^{-54}\alpha_0^2 - 2.3\times10^{-98}\alpha_0^4$ |

**Table 1:** GUP-corrections to the energy of the harmonic oscillator

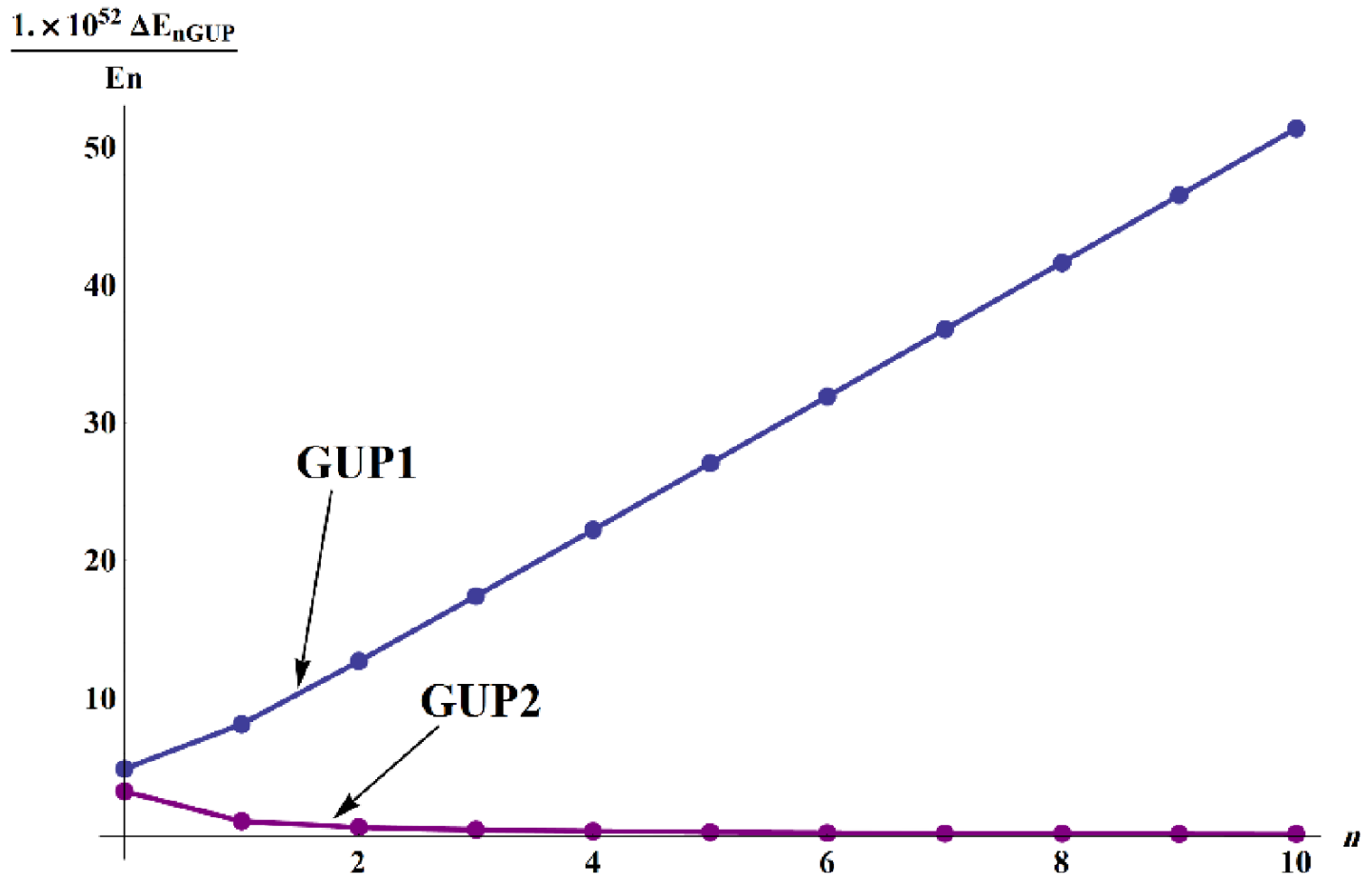

**Figure 1.** The relative change in energy due to GUP1 and GUP2 as a function of $n$, assuming $\beta_0 = \alpha_0 = 1$.

Figure 1 is a plot of (25) and (26), as a function of $n$. It is clear that the difference between the corrections of GUP1 and GUP2 increases with increasing $n$. That difference might prove useful in future experiments to differentiate between the two GUPs.

The best accuracy for mass determination for stable ions in a penning trap is [14] $\delta m / m = 1\times 10^{-11}$, which sets an upper bound on $\beta_0$ when $n = 100$ of $\beta_0 < 2.0\times 10^{38}$, and on $\alpha_0$ when $n = 1$ of $\alpha_0 < 1.7\times 10^{20}$. These bounds can be lowered in future experiments by using: Penning traps with higher mass determination accuracy, ions with higher charge, and stronger magnetic fields.

## 3. Hydrogen atom

The effect of GUP1 on the spectrum of the hydrogen atom was calculated to first order in [15] by doing the integral to find the expectation value of the perturbed Hamiltonian. In this section, we use a simpler method, adopted from [13, P.269], to get the same result. After that, we calculate the effect of GUP2 on the spectrum of hydrogen, which, to my knowledge, was not done before.

**The GUP1-corrected Hamiltonian for Hydrogen takes the form:**

$$H = \frac{p_0^2}{2m} - \frac{k}{r} + \frac{\beta}{m} p_0^4, \tag{27}$$

where $k \equiv e^2 / 4\pi\epsilon_0$, the change in energy to first order can be found as follows:

$$\Delta E_{n(GUP1)} = \langle \psi | H' | \psi \rangle = \frac{\beta}{m} \langle p_0^2 \psi | p_0^2 \psi \rangle, \tag{28}$$

where we used the hermiticity of $p_0^2 = 2m\left(E_n + \frac{k}{r}\right)$. Thus,

$$\Delta E_{n(GUP1)} = 4\beta m \left\langle \left(E_n + \frac{k}{r}\right)^2 \right\rangle = 4\beta m \left(E_n^2 + 2E_n k \left\langle \frac{1}{r} \right\rangle + k^2 \left\langle \frac{1}{r^2} \right\rangle\right). \tag{29}$$

Using the relations [13, P.269]:

$$\left\langle \frac{1}{r} \right\rangle = \frac{1}{n^2 a_0}, \qquad \left\langle \frac{1}{r^2} \right\rangle = \frac{1}{(l + 1/2) n^3 a_0^2}, \tag{30}$$

where $a_0 = 4\pi\epsilon_0 \hbar^2 / me^2 \approx 5.3\times 10^{-11} m$ is the Bohr radius, equation (29) becomes:

$$\Delta E_{n(GUP1)} = 4\beta m E_n^2 \left(1 + \frac{2k}{E_n} \frac{1}{n^2 a_0} + \frac{k^2}{E_n^2} \frac{1}{(l + 1/2) n^3 a_0^2}\right). \tag{31}$$

Using $a_0 = \hbar^2 / mk$ and $E_n = mk^2 / 2\hbar^2 n^2$, we obtain the relative change in energy:

$$\frac{\Delta E^{(1)}_{n(GUP1)}}{E_n} = 4\beta m E_n \left(\frac{4n}{l + 1/2} - 3\right), \tag{32}$$

which agrees with the expression derived in [15], and is maximum when $n = 1,\ l = 0$:

$$\frac{\Delta E^{(1)}_{1(GUP1)}}{E_1} \approx 9.3\times 10^{-49} \beta_0 . \tag{33}$$

**The GUP2-corrected Hamiltonian for Hydrogen takes the form:**

$$H = \frac{p_0^2}{2m} - \frac{k}{r} - \frac{\alpha}{m} p_0^3 + \frac{5\alpha^2}{2m} p_0^4 \tag{34}$$

The change in energy due to the $p_0^3$ term to first order is zero, because $p_0^3$ is an odd parity function, thus its integral over all space is zero.

The effect of the $p_0^4$ term is the same as (32) with $\beta \to 5\alpha^2/2$,

$$\frac{\Delta E^{(1)}_{n(GUP2)}}{E_n} = 10\alpha^2 m E_n \left( \frac{4n}{l+1/2} - 3 \right). \tag{35}$$

For $n = 1,\ l = 0$:

$$\frac{\Delta E^{(1)}_{1(GUP2)}}{E_1} \approx 2.3 \times 10^{-48} \alpha_0^2 . \tag{36}$$

The second order correction for the $p_0^3$ term, can be found numerically, for the ground state $\psi_{100}$:

$$\Delta E^{(2)}_{1(GUP2)\, p_0^3} = \sum_{nlm \neq 100}^{\infty} \frac{\left| \langle nlm | H' | 100 \rangle \right|^2}{E_1^{(0)} - E_n^{(0)}}, \qquad H' = \frac{-\alpha}{m} p_0^3 = \frac{\alpha i \hbar^3}{m} \nabla\left(\nabla^2\right), \tag{37}$$

From selection rules [13, P.360] $\langle nlm | p | n'l'm' \rangle = 0$ except when $\Delta m = \pm 1, 0$ and $\Delta l = \pm 1$, which means that the sum should be taken for $l = 1,\ m = -1, 0, 1$. Summing for all states adjacent to $|100\rangle$ (e.g. up to $n = 10$), since their contribution is greater:

$$\Delta E^{(2)}_{1(GUP2)} = \frac{\alpha^2 \hbar^6}{m^2 E_1} \sum_{\substack{n=2, l=1,\\ m=0,\pm 1}}^{n=10} \frac{1}{1 - \frac{1}{n^2}} \left| \int_0^{2\pi} \int_0^{\pi} \int_0^{\infty} \psi_{nlm} \nabla\nabla^2 \left(\psi_{100}\right) r^2 \sin\theta dr d\theta d\phi \right|^2 \tag{38}$$

The gradient of the Laplacian of $\psi_{100}$ in spherical coordinates is:

$$\nabla\nabla^2\left(\psi_{100}\right) = \frac{1}{\sqrt{\pi} r^2} \left( \frac{1}{a_0^3} \right)^{3/2} e^{-\frac{r}{a_0}} \left( 2a_0^2 + 2a_0 r - r^2 \right) \hat{\mathbf{r}} \tag{39}$$

Substituting in (38) taking into consideration that $\hat{\mathbf{r}} = \sin\theta\cos\phi \hat{\mathbf{x}} + \sin\theta\sin\phi \hat{\mathbf{y}} + \cos\theta \hat{\mathbf{z}}$, leads to:

$$\frac{\Delta E^{(2)}_{1(GUP2)}}{E_1} \simeq 6.2 \times 10^{-52} \alpha_0^2 \tag{40}$$

which is much less than (36), and thus can be neglected; this also happens to all other states.

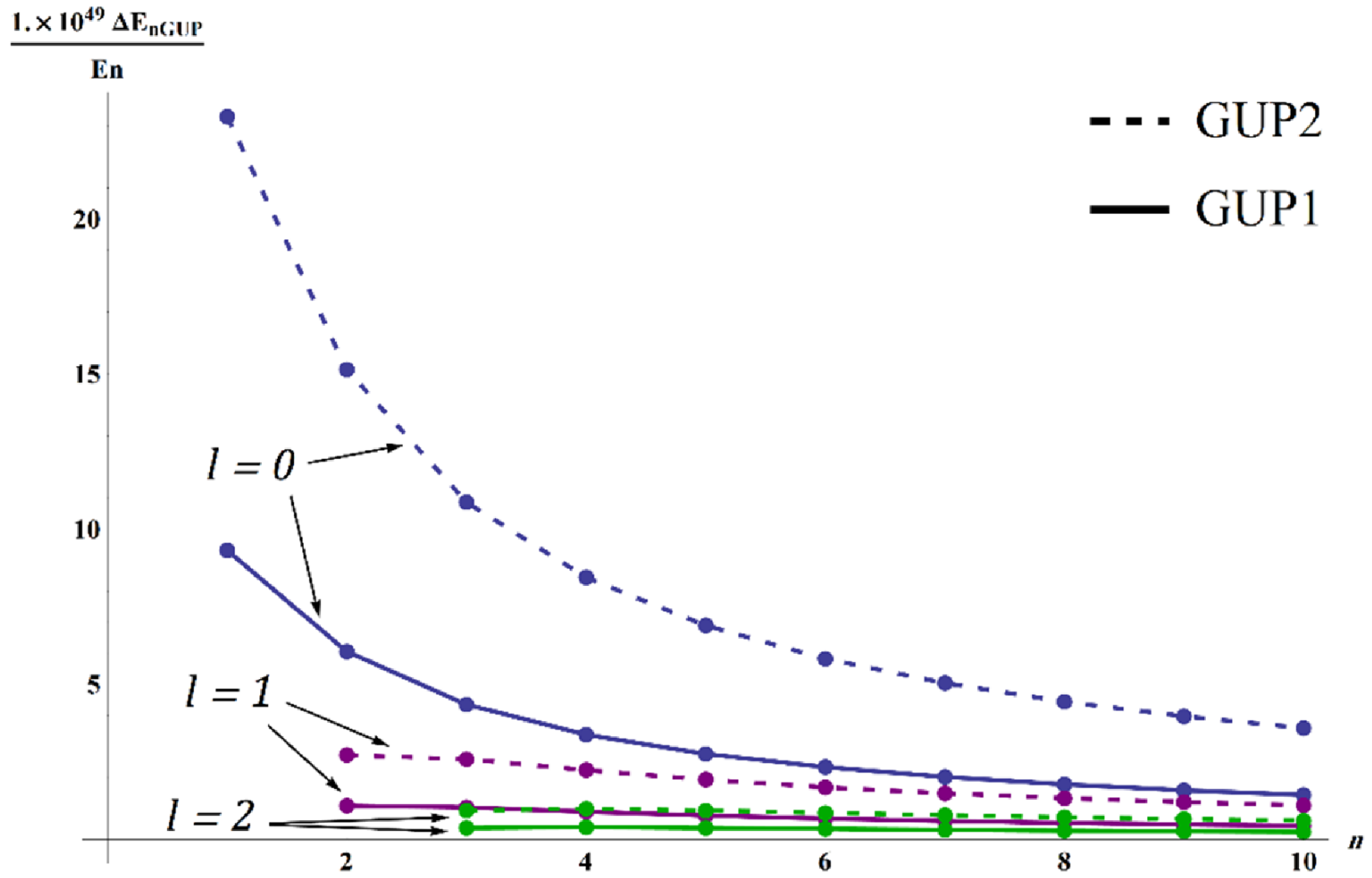

**Figure 2:** GUP-corrections to the spectrum of the Hydrogen atom.

Figure 2 is a plot of (32) and (35) as a function of $n$ for different $l$; we see that the two GUPs have almost the same effect on the spectrum of Hydrogen. The best experimental measurement of the 1S-2S transition in Hydrogen [16] reaches a fractional frequency uncertainty of $\delta f / f = 4.2 \times 10^{-15}$ which sets an upper bound on $\beta_0$ of $\beta_0 < 4.5 \times 10^{33}$, and on $\alpha_0$ of $\alpha_0 < 4.2 \times 10^{16}$.

## 4. Modified Lorentz force law

Because the GUP modifies the Hamiltonian, one expects that any system with a well-defined Hamiltonian is perturbed [9], perhaps even classical Hamiltonians. The impact of the GUP2-corrected classical Hamiltonian on Newton's gravitational force law was examined in [17]; here, we derive a modified Lorentz force law.

For a particle in an electromagnetic field, the GUP1-modified Hamiltonian is [5]:

$$H = \frac{1}{2m}(\boldsymbol{p}_0 - q\boldsymbol{A})^2 + \frac{\beta}{m}(\boldsymbol{p}_0 - q\boldsymbol{A})^4 + q\varphi, \tag{41}$$

differentiating w.r.t $\boldsymbol{p}_0$ :

$$\dot{\boldsymbol{r}} = \frac{\partial H}{\partial \boldsymbol{p}_0} = \frac{1}{m}(\boldsymbol{p}_0 - q\boldsymbol{A}) + \frac{4\beta}{m}(\boldsymbol{p}_0 - q\boldsymbol{A})^3 \tag{42}$$

Using inversion of series:

$$\boldsymbol{p}_0 = q\boldsymbol{A} + m\dot{\boldsymbol{r}} - 4\beta\left(m\dot{\boldsymbol{r}}\right)^3 + O\left(\beta^2\right) \tag{43}$$

Substitution in $\mathcal{L} = \boldsymbol{p}_0.\dot{\boldsymbol{r}} - H$ leads to:

$$\mathcal{L} = \left(m\dot{\boldsymbol{r}} - 4\beta\left(m\dot{\boldsymbol{r}}\right)^3 + q\boldsymbol{A}\right).\dot{\boldsymbol{r}} - \frac{1}{2m}\left(m\dot{\boldsymbol{r}} - 4\beta\left(m\dot{\boldsymbol{r}}\right)^3\right)^2 - \frac{\beta}{m}\left(m\dot{\boldsymbol{r}} - 4\beta\left(m\dot{\boldsymbol{r}}\right)^3\right)^4 - q\varphi. \tag{44}$$

Simplifying:

$$\mathcal{L} = \frac{m\dot{\boldsymbol{r}}^2}{2} - \beta m^3\dot{\boldsymbol{r}}^4 + q\boldsymbol{A}.\dot{\boldsymbol{r}} - q\varphi . \tag{45}$$

Applying the Euler-Lagrange equation $\frac{d}{dt}\left(\frac{\partial \mathcal{L}}{\partial \dot{r}}\right) - \frac{\partial \mathcal{L}}{\partial r} = 0$ we obtain:

$$m\ddot{\boldsymbol{r}} - 12\beta m^3\dot{\boldsymbol{r}}^2\ddot{\boldsymbol{r}} = q\nabla\left(\boldsymbol{A}.\dot{\boldsymbol{r}}\right) - q\frac{d\boldsymbol{A}}{dt} - q\nabla\phi . \tag{46}$$

The RHS is $q(\boldsymbol{E} + \boldsymbol{v} \times \mathbf{B})$, which means that the Lorentz force law becomes:

$$\boldsymbol{F} \equiv m\ddot{\boldsymbol{r}} = q\frac{\boldsymbol{E} + \boldsymbol{v} \times \mathbf{B}}{1 - 12\beta m^2 v^2}, \tag{47}$$

which is approximately:

$$\boldsymbol{F} \simeq q\left(\boldsymbol{E} + \boldsymbol{v} \times \mathbf{B}\right)\left(1 + 12\beta m^2 v^2\right). \tag{48}$$

Using the same method as above, the GUP2-corrected Hamiltonian takes the form [8]:

$$H = \frac{1}{2m}(\boldsymbol{p}_0 - q\boldsymbol{A})^2 - \frac{\alpha}{m}(\boldsymbol{p}_0 - q\boldsymbol{A})^3 + \frac{5\alpha^2}{2m}(\boldsymbol{p}_0 - q\boldsymbol{A})^4 + q\varphi, \tag{49}$$

differentiating w.r.t $\boldsymbol{p}_0$ and using inversion of series:

$$\boldsymbol{p}_0 = q\boldsymbol{A} + m\dot{\boldsymbol{r}} + 3\alpha\left(m\dot{\boldsymbol{r}}\right)^2 + 8\alpha^2\left(m\dot{\boldsymbol{r}}\right)^3 + O\left(\alpha^3\right), \tag{50}$$

leading to the Lagrangian:

$$\mathcal{L} = \frac{m\dot{\boldsymbol{r}}^2}{2} + \alpha m^2\dot{\boldsymbol{r}}^3 + 2\alpha^2 m^3\dot{\boldsymbol{r}}^4 + q\boldsymbol{A}.\dot{\boldsymbol{r}} - q\varphi . \tag{51}$$

from which we obtain:

$$\boldsymbol{F} = q\frac{\boldsymbol{E} + \boldsymbol{v} \times \mathbf{B}}{1 + 6\alpha m\boldsymbol{v} + 24\alpha^2 m^2 v^2}, \tag{52}$$

which is approximately:

$$\boldsymbol{F} \simeq q\left(\boldsymbol{E} + \boldsymbol{v} \times \mathbf{B}\right)\left(1 - 6\alpha m v\right). \tag{53}$$

The new term in (48) and (53) depends on $mv$ , which means that its effect in high energy physics will be too small even at relativistic speeds. For example, in a proton-proton scattering experiment:

$$\Delta F_{GUP1} / F = 12\beta m^2 v^2 \sim 10^{-38}\beta_0 . \tag{54}$$

Experimental tests of Coulomb’s law use large, but usually static, masses [18]. For example, coulomb’s torsion balance experiment measures the torsion force needed to balance the electrostatic force, and Cavendish’s concentric spheres experiment, and its modern counterparts, use two or more concentric spheres, (or cubes, or icosahedra) [18] to test Gauss’s law.

To test Eqs. (48) and (53) we need large masses, with moderate velocities. Suppose we have a pendulum with length $R$ and a bob with charge $q$ and mass $m$ swinging above an infinite charged plane with charge density $-\sigma$ ; the electric field will be $E = -\sigma / 2\varepsilon_0$ . Without the GUP effect, the bob will experience a force:

$$\boldsymbol{F_0} = -\left( mg + \frac{q\sigma}{2\varepsilon_0} \right)\hat{\boldsymbol{y}} . \tag{55}$$

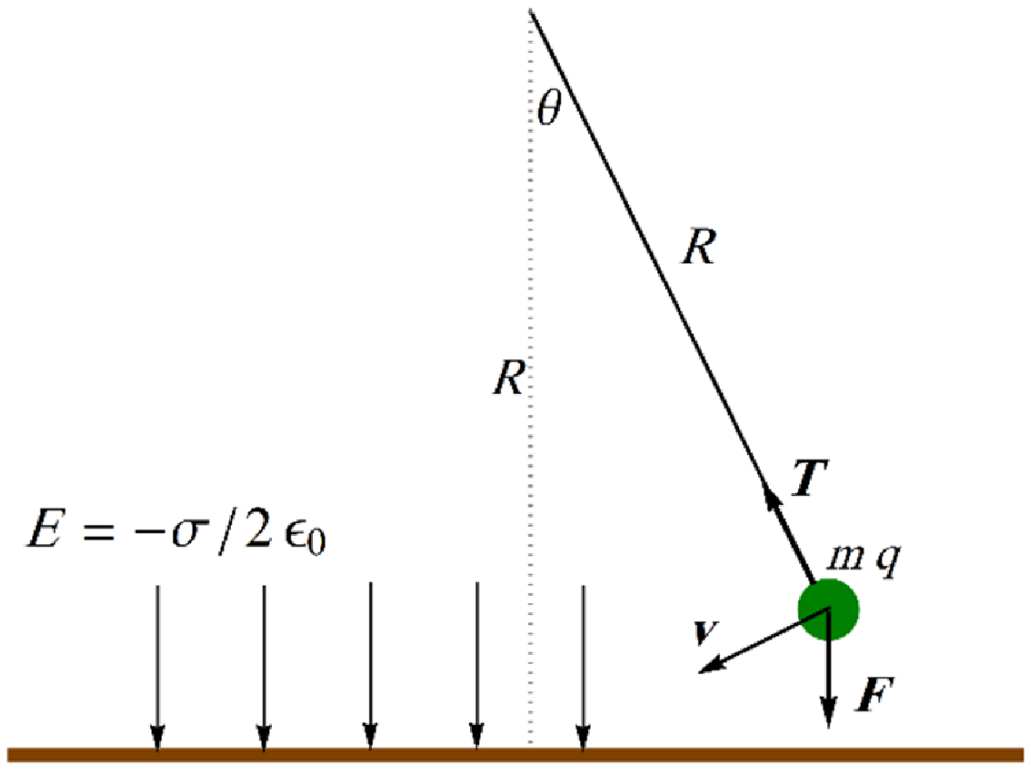

**Figure 3:** A pendulum under the effect of gravitational and electrostatic forces.

If $\theta$ is the angle between the vertical and the string, the equation of motion for small $\theta$ is:

$$mR\ddot{\theta} \simeq -\left( mg + \frac{q\sigma}{2\varepsilon_0} \right)\theta . \tag{56}$$

Thus, the angular frequency is:

$$\omega_0^{\ 2} = \frac{g}{R} + \frac{q\sigma}{2\varepsilon_0 mR} . \tag{57}$$

However, if we used Eq. (48) for the electrostatic force, then the equation of motion will be:

$$mR\ddot{\theta} \simeq -\left( mg + \frac{q\sigma}{2\varepsilon_0}\left(1 + 12\beta m^2 v^2\right) \right)\theta . \tag{58}$$

The velocity can be found from conservation of energy, taking the gravitational and electrical potentials to be zero on the plane:

$$\frac{1}{2}mv^2 + \left( \frac{\sigma q}{2\varepsilon_0} + mg \right)R\left(1 - \cos\theta\right) = \left( \frac{\sigma q}{2\varepsilon_0} + mg \right)R\left(1 - \cos\theta_0\right), \tag{59}$$

where $\theta_0$ is the initial angle, assuming it starts with zero initial velocity.

$$v^2 \simeq \left( \frac{\sigma q}{2m\varepsilon_0} + g \right)R\left(\theta_0^2 - \theta^2\right). \tag{60}$$

The equation of motion will be:

$$mR\ddot{\theta} \simeq -\left( mg + \frac{q\sigma}{2\varepsilon_0} + \frac{6q\sigma\beta m^2}{\varepsilon_0}\left( \frac{\sigma q}{2m\varepsilon_0} + g \right)R\theta_0^{\ 2} \right)\theta . \tag{61}$$

Thus, the angular frequency is:

$$\omega_1^{\ 2} = \frac{g}{R} + \frac{q\sigma}{2mR\varepsilon_0} + \frac{6q\sigma\beta}{\varepsilon_0}\left( \frac{\sigma q}{2\varepsilon_0} + mg \right)\theta_0^{\ 2} . \tag{62}$$

And for GUP2

$$\omega_2^{\ 2} = \frac{g}{R} + \frac{q\sigma}{2mR\varepsilon_0} - \frac{3q\sigma\alpha}{\varepsilon_0}\theta_0\sqrt{\frac{\sigma q}{2mR\varepsilon_0} + \frac{g}{R}} . \tag{63}$$

Using the values $\theta_0 = \pi/12,\ \sigma = 1\mu C/m^2,\ q = 2\mu C,\ R = 1m,\ m = 0.1kg, g = 9.807m/s^2$ :

$$\omega_0 = 3.307 rad/\sec, \quad F_0 = 1.094N , \tag{64}$$

$$\omega_1 = 3.307 + 3.6\times10^{-4}\beta_0, \quad F_1 = 1.094 + 2.3\times10^{-3}\beta_0, \tag{65}$$

$$\omega_2 = 3.307 - 1.4\times10^{-2}\alpha_0, \quad F_2 = 1.094 - 8.6\times10^{-2}\alpha_0 . \tag{66}$$

These values, I believe, are accessible with current technology, and thus can be used to set much lower bounds on the GUP parameters than the best bound [19] of $\alpha_0 < 10^8$ from the anomalous magnetic moment of the muon. However, the GUP might not be applicable on large scale; maybe the GUP parameters $\alpha_0$ and $\beta_0$ are mass dependent.

## 5. Generalized Time-Energy Uncertainty

Suppose a light-clock consists of two parallel mirrors a distance $L$ apart, the time a photon takes to travel from one mirror to the other is: $T = L/c$ , but length cannot be measured more accurately than the Planck length so:

$$T = \frac{L \pm l_P}{c} = \frac{L}{c} \pm t_P, \tag{67}$$

where $t_P \equiv \sqrt{G\hbar/c^5} \simeq 5.4\times10^{-44} sec$ , is the Planck time. This shows that the existence of a minimal length scale limits the precision of time measurements. A more rigorous analysis using general relativity, and taking into account the gravitational attraction between the photon and the mirrors, leads to the same conclusion [1, 20].

The time-energy uncertainty relation can be obtained from the position-momentum uncertainty relation by using $p = E/c$ and $t = x/c$ to give:

$$\Delta E\,\Delta \mathrm{t} \ge \hbar/2 \tag{68}$$

GUP1 leads to the generalized time-energy uncertainty relation:

$$\Delta E\,\Delta \mathrm{t} \ge \frac{\hbar}{2}\left[1+3\frac{\beta}{c^2}\left((\Delta E)^2+\langle E\rangle^2\right)\right], \tag{69}$$

which implies $\Delta t \ge \Delta t_{min} = \sqrt{\beta_0}\, t_P$ . GUP2 leads to:

$$\Delta E\,\Delta \mathrm{t} \ge \frac{\hbar}{2}\left[1-2\frac{\alpha}{c}\langle E\rangle+4\frac{\alpha^2}{c^2}\left((\Delta E)^2+\langle E\rangle^2\right)\right], \tag{70}$$

which implies $\Delta t \ge \Delta t_{min} = \alpha_0 t_P$ .

An important application of the time-energy uncertainty is calculating the mean life $\tau$ of short-lived particles, by using the full width $\Gamma$ divided by two as a measure of $\Delta E$ [21], i.e. $\tau = \hbar/\Gamma$ , because $\Gamma$ is easier to determine experimentally than $\tau$ . Applying (69) & (70) instead of (68) leads to an extremely small change in the mean life of particles.

In table 2, the mass $m$ and the full width $\Gamma$ are from [22]. The mean life was calculated via (68), while $\Delta\tau_{GUP1}$ and $\Delta\tau_{GUP2}$ were calculated via (69) and (70) respectively. The rest mass was used as a measure of $\langle E\rangle$ .

| Particle | Mass $m$ [MeV] | Full width $\Gamma$ [MeV] | Mean life $\tau$ [sec.] | $\Delta\tau_{GUP1}/\tau$ | $\Delta\tau_{GUP2}/\tau$ |
|---|---|---|---|---|---|
| $Z$ | $91.19\times10^3$ | $2.49\times10^3$ | $2.64\times10^{-25}$ | $1.7\times10^{-34}\beta_0$ | $-2.3\times10^{-17}\alpha_0$ |
| $\eta$ | 547.85 | $1.30\times10^{-3}$ | $5.06\times10^{-19}$ | $6.1\times10^{-39}\beta_0$ | $-8.9\times10^{-20}\alpha_0$ |
| $\mu$ | 105.66 | $2.99\times10^{-16}$ | $2.197\times10^{-6}$ | $2.2\times10^{-40}\beta_0$ | $-1.7\times10^{-20}\alpha_0$ |

**Table 2:** Effect of the modified time-energy uncertainty principle on the mean life of particles

The effect of the generalized time-energy uncertainty principle on the mean life is too small to measure experimentally, but it might affect the Planck era cosmology [23]. In [23] the authors investigate the effect of similar relations to (69) & (70) on the values of the main Planck quantities, like $t_P$, and reach the conclusion that they were lager at the Planck era than now by a factor of $(10-10^4)$ under specific conditions. If true, then the effect of (69) & (70) on the mean life of particles was greater at the early universe, and might leave traces in present day cosmology.

## 6. Conclusions

In this paper, we investigated some implications of the GUP1 and GUP2. We calculated the GUP-corrections to the energy of the quantum harmonic oscillator for all energy levels to first and second order perturbation; and although the corrections are small, current and future experiments can be used to set bounds on the values of the GUP parameters. We also found that the difference between corrections due to GUP1 and GUP2 gets bigger with increasing $n$ ; this may provide a way to experimentally determine which GUP is correct.

Then, we investigated the GUP-effect on the spectrum of atomic Hydrogen, because spectroscopy provides increasingly more precise measurements for transition frequencies in atoms. We also found that GUP1 and GUP2 have almost the same effect on the spectrum of Hydrogen.

After that, we investigated how the GUP-corrected classical Hamiltonian leads to a modified Lorentz force law. We also found that it might be possible to detect the effect of the modified Lorentz force law with current technology, unless the GUP is only applicable near the Planck scale.

Finally, we saw how the GUP leads to a generalized time-energy uncertainty principle, and considered it effect on the mean life of some particles, which was too small to measure experimentally. However, its effect in the early universe might be detectable in present day cosmology.

## 7. Acknowledgements

I would like to thank Dr. Ahmed Farag Ali for his support, and for the interesting discussions we had on the GUP. I would also like to thank the anonymous referees whose useful comments and suggestions made this paper much better.

## 8. References

[1] Hossenfelder, S. (2013). Minimal length scale scenarios for quantum gravity. [arXiv:1203.6191], 86. 'and references therein'

[2] Ng, Y. J. (2003). Selected topics in Planck-scale physics. Modern Physics Letters A, 18(16), 1073-1097. [arXiv: gr-qc/0305019] 'and references therein'

[3] Maggiore, M. (1993). A generalized uncertainty principle in quantum gravity. Physics Letters B, 304(1), 65-69. [arXiv: hep-th/9301067]

[4] Garay, L. J. (1995). Quantum gravity and minimum length. International Journal of Modern Physics A, 10(02), 145-165. [arXiv: gr-qc/9403008]

[5] Das, S., & Vagenas, E. C. (2009). Phenomenological Implications of the Generalized Uncertainty Principle. [arXiv:0901.1768]

[6] Kempf, A., Mangano, G., & Mann, R. B. (1994). Hilbert space representation of the minimal length uncertainty relation. [arXiv: hep-th/9412167].

[7] Ali, A. F., Das, S., & Vagenas, E. C. (2009). Discreteness of space from the generalized uncertainty principle. Physics Letters B, 678(5), 497-499. [arXiv: 0906.5396]

[8] Ali, A. F., Das, S., & Vagenas, E. C. (2011). Proposal for testing quantum gravity in the lab. Physical Review D, 84(4), 044013. [arXiv:1107.3164]

[9] Das, S., & Vagenas, E. C. (2008). Universality of quantum gravity corrections. Physical review letters, 101(22), 221301. [arXiv:0810.5333]

[10] Kempf, A. (1997). Non-pointlike particles in harmonic oscillators. Journal of Physics A: Mathematical and General, 30(6), 2093. [arXiv: hep-th/9604045]

[11] Pikovski, I., Vanner, M. R., Aspelmeyer, M., Kim, M. S., & Brukner, Č. (2012). Probing Planck-scale physics with quantum optics. Nature Physics, 8(5), 393-397. [arXiv: 1111.1979]

[12] Chang, L. N., Minic, D., Okamura, N., & Takeuchi, T. (2002). Exact solution of the harmonic oscillator in arbitrary dimensions with minimal length uncertainty relations. Physical Review D, 65(12), 125027. [arXiv: hep-th/0111181]

[13] Griffiths, D. J. (2005). Introduction to Quantum Mechanics, 2/E. Pearson Education.

[14] Blaum, K. (2006). High-accuracy mass spectrometry with stored ions. Physics Reports, 425(1), 1-78.

[15] Brau, F. (1999). Minimal length uncertainty relation and the hydrogen atom. Journal of Physics A: Mathematical and General, 32(44), 7691. [arXiv: quant-ph/9905033]

[16] Parthey, C. G., et al. (2011). Improved Measurement of the Hydrogen 1S–2S Transition Frequency. Physical Review Letters, 107(20), 203001. [arXiv:1107.3101]

[17] Ali, A. F. (2011). Minimal length in quantum gravity, equivalence principle and holographic entropy bound. Classical and Quantum Gravity, 28(6), 065013. [arXiv:1101.4181]

[18] Tu, L. C., & Luo, J. (2004). Experimental tests of Coulomb's Law and the photon rest mass. Metrologia, 41(5), S136.

[19] Das, S., & Mann, R. B. (2011). Planck scale effects on some low energy quantum phenomena. Physics Letters B, 704(5), 596-599. [arXiv: 1109.3258].

[20] Padmanabhan, T. (1987). Limitations on the operational definition of spacetime events and quantum gravity. Classical and Quantum Gravity, 4(4), L107.

[21] Griffiths, D. (2008). Introduction to elementary particles. Wiley-Vch. P.198

[22] Particle Data Group. (2012). Particle physics booklet. Institute of Physics publishing.

[23] Basilakos, S., Das, S., & Vagenas, E. C. (2010). Quantum Gravity corrections and entropy at the Planck time. Journal of Cosmology and Astroparticle Physics, 2010(09), 027. [ arXiv:1009.0365]